\documentclass[prl,aps,lengthcheck,twocolumn,nofootinbib,notitlepage,floatfix,superscriptaddress]{revtex4-2}

\usepackage{graphics,graphicx}
\usepackage{amsmath, amssymb}
\usepackage{multirow}
\usepackage{color}
\usepackage{verbatim}
\usepackage{hyperref}
\usepackage[normalem]{ulem}
\usepackage{color}
\usepackage{cancel}
\usepackage{mathtools}
\usepackage{orcidlink}

\def\fm3{\,\text{fm}^{-3}}

\def \scaleFig {1}

\renewcommand\sout{\bgroup\color{blue} \ULdepth=-.5ex \ULset}
\def\slashchar#1{\setbox0=\hbox{$#1$}  
\dimen0=\wd0     
\setbox1=\hbox{/} \dimen1=\wd1  
\ifdim\dimen0>\dimen1   
\rlap{\hbox to \dimen0{\hfil/\hfil}} 
#1     
\else     
\rlap{\hbox to \dimen1{\hfil$#1$\hfil}} 
/      
\fi}

\newcommand{\dd}{\mathrm{d}}
\newcommand{\pp}{\partial}

\allowdisplaybreaks

\begin{document}

\title{Hadron spectra and thermodynamics for all quark flavors from a universal Hagedorn temperature}

\author{Micha\l{} Marczenko \orcidlink{0000-0003-2815-0564}}
\email{michal.marczenko@uwr.edu.pl}
\affiliation{Institute of Theoretical Physics, University of Wroc\l{}aw, plac Maksa Borna 9, PL-50204 Wroc\l{}aw, Poland}
\author{Larry McLerran \orcidlink{0000-0002-0733-3110}}
\affiliation{Institute for Nuclear Theory, University of Washington, Box 351550, Seattle, WA 98195, USA}
\author{Krzysztof Redlich \orcidlink{0000-0002-2629-1710}}
\affiliation{Institute of Theoretical Physics, University of Wroc\l{}aw, plac Maksa Borna 9, PL-50204 Wroc\l{}aw, Poland}
\affiliation{Polish Academy of Sciences PAN, Podwale 75, 
PL-50449 Wroc\l{}aw, Poland}

\date{\today}

\begin{abstract}
We show that hadrons in QCD follow a spectrum determined by string dynamics characterized by a universal Hagedorn temperature linked to the string tension. While this behavior was recently established for light hadrons and glueballs, we demonstrate that the same dynamics describes the heavy-flavor sector. After separating the current quark masses, the resulting spectrum reproduces lattice QCD thermodynamics of charmed hadrons and the observed spectra of hadrons across quark flavors without additional parameters. These results reflect the universal confining dynamics of QCD through the string tension.
\end{abstract}

\maketitle

\emph{Introduction.---}
Quantum Chromodynamics (QCD) predicts that strongly interacting matter undergoes a transition from a confined hadronic phase to a deconfined quark-gluon plasma at sufficiently high temperature\,\cite{Gross:2022hyw}. Below the crossover temperature \mbox{$T_c = 156.5\pm1.5\,$MeV\,\cite{HotQCD:2018pds}}, thermodynamics is dominated by hadronic degrees of freedom whose spectrum strongly influences the equation of state in the confined phase\,\cite{Braun-Munzinger:2003pwq, Andronic:2017pug, Aarts:2023vsf}.

Lattice QCD (LQCD) provides first-principles calculations of hadron masses, but present simulations determine only ground states and a limited number of low-lying excitations for fixed quantum numbers. However, finite-temperature LQCD calculations below $T_c$ are governed by the complete hadronic density of states\,\cite{Majumder:2010ik}. As a consequence, thermodynamic observables such as the equation of state and fluctuations of conserved charges are driven by the proliferation of hadronic excitations\,\cite{Bazavov:2014xya, Bazavov:2014pvz, Borsanyi:2018grb, Bellwied:2019pxh, Bollweg:2021vqf}.

The rapid growth of states arises naturally in a string description of confinement and hadron spectroscopy. In gauge theory, the long-distance interaction between static color sources is characterized by a linear potential corresponding to a color flux tube with tension $\sigma$. This flux tube admits an effective description as a relativistic open string whose endpoints carry quark degrees of freedom\,\cite{Nambu:1974zg, Isgur:1984bm}. Quantization of such a string leads to an exponentially increasing number of vibrational modes with mass, giving rise to a Hagedorn spectrum governed by a limiting temperature $T_H$\,\cite{Green_Schwarz_Witten_2012, Hagedorn:1965st, Hagedorn:1971mc}. In this picture, mesons correspond to open strings with quarks attached to their endpoints\,\cite{Fujimoto:2025sxx, Marczenko:2025nhj}, while baryons can be described effectively by a quark-diquark configuration connected by a single flux tube\,\cite{Selem:2006nd, Fujimoto:2025trl}. On the other hand, glueballs can be interpreted as closed strings without endpoints~\cite{Green_Schwarz_Witten_2012}. Lattice studies of the glueball spectrum are also consistent with an exponential growth\,\cite{Miller:2006hr, Athenodorou:2020ani}.

It has recently been shown that the lattice equation of state and fluctuations of conserved charges in $N_f=2+1$ QCD, as well as thermodynamics in pure gauge theory, are quantitatively described in the confined phase by a continuous density of states characterized by a limiting temperature $T_H \simeq 300$-$340$ MeV\,\cite{Fujimoto:2025sxx, Marczenko:2025nhj, Fujimoto:2025trl, Marczenko:2026yme}, consistent with the observed exponential growth of the hadronic density of states. This temperature is proportional to the square root of the string tension, $T_H/\sqrt{\sigma}=\sqrt{3/2\pi}$; thus, $T_H$ is interpreted as an effective scale set by the universal string tension. Recent lattice determinations of the string tension, $\sqrt\sigma = 0.485(6)\,$GeV\,\cite{Athenodorou:2020ani} and $\sqrt\sigma = 0.4817(97)\,$GeV\,\cite{Brambilla:2022het}, correspond to $T_H = 0.335(4)\,$GeV and $T_H = 0.3329(67)\,$GeV, respectively. Heavy quarks introduce large additional mass scales, and it is therefore not obvious that the same Hagedorn temperature should govern the excitation spectrum of heavy-flavor hadrons (see Ref.\,\cite{Chen:2022asf} for a recent review of heavy-flavor spectroscopy). Whether this temperature persists across quark flavors, including hadrons containing heavy quarks, remains unclear. It was recently conjectured that the Hagedorn temperature might even be linked to the thermal dilepton production rate\,\cite{Stoecker:2026jyy}.

In this work, we demonstrate that the Hagedorn spectrum is universal across all quark flavors. The key observation is that the exponential growth of states is controlled by the excitation energy of the confining flux tube. We show that isolating the non-dynamical, current quark masses leads to a universal description of hadronic spectra with a single Hagedorn temperature. Using a string-based density of states with $T_H$ fixed entirely from the light sector, we confront thermodynamic observables and spectral abundances in the charm and bottom sectors. We find that the same Hagedorn temperature consistently describes open- and hidden-heavy hadrons without additional free parameters once the current quark masses are separated. 
This demonstrates that the proliferation of hadronic states is governed by a universal mass spectrum of open strings and a scale set by the QCD string tension, independent of quark flavor.

\emph{Thermodynamics of charm.---}
Thermodynamic observables are determined by the density of states through the grand-canonical pressure. In the Boltzmann approximation, the pressure of a hadronic state of mass $m$ carrying baryon number $B$, electric charge $Q$, strangeness $S$, and charm $C$ reads
\begin{equation}
\hat P(T,\mu,m)
=
\frac{1}{2\pi^2}
\left(\frac{m}{T}\right)^2
K_2\!\left(\frac{m}{T}\right)
e^{\mu/T},
\end{equation}
where
$
\mu=B\mu_B+Q\mu_Q+S\mu_S+C\mu_C,
$
with $\mu_i$ denoting the chemical potential associated with the corresponding quantum number. The total pressure follows from integrating over the density of states $\rho(m)$,
\begin{equation}\label{eq:pres}
\hat P(T,\mu) = \int\limits_0^\infty dm \, \rho(m)\, \hat P(T,\mu,m).
\end{equation}

\begin{figure}[t!]
    \centering
    \includegraphics[width=\scaleFig\linewidth]{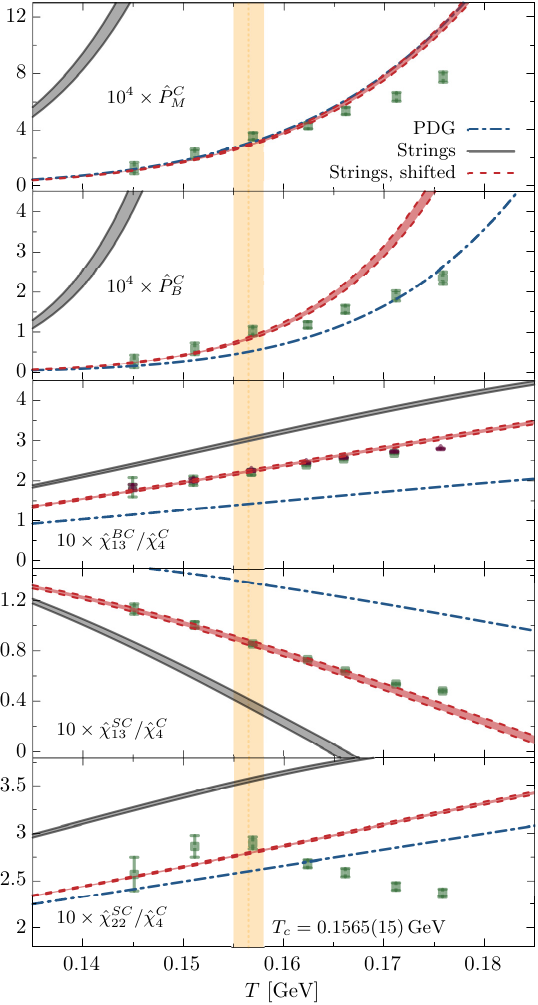}
    \caption{
    Partial pressures of open-charm mesons and charmed baryons, together with the ratios $\hat \chi_{13}^{BC}/\hat \chi_4^C$, $\hat \chi_{13}^{SC}/\hat \chi_4^C$, and $\hat \chi_{22}^{SC}/\hat \chi_4^C$ (top to bottom) as functions of temperature. Results obtained from the Hagedorn spectrum without quark-mass subtraction (solid black lines) and with subtraction (red dashed lines) are compared to lattice QCD calculations~\cite{Kaczmarek:2025dqt} (points) and to the hadron resonance gas (HRG) model with PDG input (blue dash-dotted lines)~\cite{Kaczmarek:2025dqt}. For the pressures, lattice data correspond to continuum estimates~\cite{Kaczmarek:2025dqt} and for susceptibilities to $N_\tau = 8$ simulations along lines of constant physics~\cite{Kaczmarek:2025dqt, Bazavov:2023xzm}. Shaded bands reflect the uncertainty of the Hagedorn temperature, $T_H = 323(3)\,$MeV~\cite{Marczenko:2026yme}. The vertical yellow band indicates the chiral crossover temperature $T_c = 156.5 \pm 1.5\,$MeV~\cite{HotQCD:2018pds}.
    }
    \label{fig:p_x_ratios}
\end{figure}

In gauge theory, the interaction between static color sources is characterized at large distances by a linear potential, $V(r)\simeq \sigma r$, where $\sigma$ is the string tension. This suggests an effective description of hadrons as open relativistic flux tubes with quarks attached to their endpoints. Quantization of such strings leads to a rapidly increasing number of vibrational modes with mass, and the asymptotic density of states takes a universal form
\begin{equation}\label{eq:rho_string}
    \rho_{\rm str}(m) =
    \frac{\sqrt{2\pi}}{6T_H}\left(\frac{m}{T_H}\right)^{-3/2}e^{m/T_H},
\end{equation}
where $T_H$ denotes the Hagedorn temperature, related to the string tension $\sigma = 2T_H^2/3\pi$ for a bosonic string in four spacetime dimensions\,\cite{Green_Schwarz_Witten_2012}.

The observed hadron spectra can be represented by introducing a set of hadronic mass thresholds,
\begin{equation}\label{eq:rho_thresholds}
    \rho(m) = \sum_i g_i \Theta(m-m^i_{\rm thr})\rho_{\rm str}(m),
\end{equation}
corresponding to the lightest state in each sector, with $m^i_{\rm thr}$ and $g_i$ denoting their masses and degeneracy factors. The index $i$ runs over the mass thresholds considered in the spectrum. The construction of the hadronic thresholds entering Eq.~\eqref{eq:rho_thresholds} is discussed in End Matter.

\begin{figure*}
    \centering    \includegraphics[width=\scaleFig\linewidth]{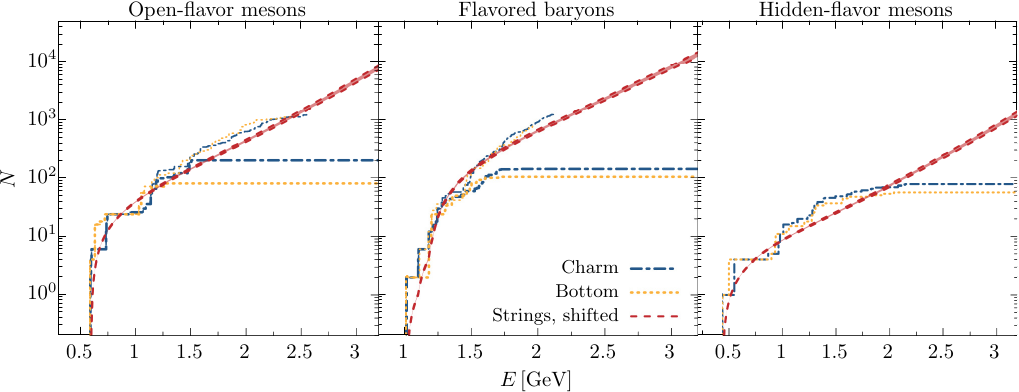}
     \caption{Cumulative spectra of heavy-flavor hadrons as functions of the excitation energy $E,$ as in Eq.\,\eqref{eq:Ex}. Shown are open-flavor mesons (left panel), singly-heavy baryons (middle panel), and hidden-flavor mesons (right panel). Thick dotted and dash-dotted lines denote experimentally established mesons or baryons with three- and four-star rating in the Particle Data Group (PDG)\,\cite{Kaczmarek:2025dqt, ParticleDataGroup:2024cfk}. Thin dotted and dash-dotted lines show spectra predicted by the quark model (QM)\,\cite{Ebert:2009ua, Ebert:2011kk}, shown only up to the mass where calculations are available in all channels. Red dashed curves represent the universal Hagedorn spectrum with $T_H=323(3)\,$MeV\,\cite{Marczenko:2026yme}, constructed using the same excitation-energy threshold determined in the charm sector without additional adjustment (see text for details). 
     The shaded bands reflect the uncertainty in $T_H$.}
    \label{fig:spectra_bottom}
\end{figure*}

The relation between the Hagedorn spectrum and hadronic spectroscopy can be understood within the effective string description of confinement. In this framework, the linear Regge trajectories observed in meson and baryon spectroscopy arise from the classical rotation of a relativistic flux tube with tension $\sigma$, yielding $J\simeq \alpha' M^2$ with $\alpha'=(2\pi\sigma)^{-1}$\,\cite{Chew:1962eu, Selem:2006nd, Sonnenschein:2014jwa}. For hadrons containing heavy quarks, endpoint masses modify the classical string dynamics and lead to effective Regge slopes that differ from those of light hadrons\,\cite{Sonnenschein:2014jwa, Sonnenschein:2018fph}. The exponential growth of states, in contrast, originates from the quantized vibrational modes of the same flux tube. Both phenomena therefore probe the same underlying string dynamics but correspond to different limits: Regge trajectories reflect classical rotational motion, whereas the Hagedorn spectrum arises from the multiplicity of quantum excitations\,\cite{Nambu:1974zg, Green_Schwarz_Witten_2012}. In realistic QCD systems, additional effects, including endpoint masses, Coulomb interactions between heavy quarks, and hyperfine spin splittings, modify the lowest-lying states and the slopes of individual Regge trajectories\,\cite{Sonnenschein:2014jwa, Sonnenschein:2018fph, Kruczenski:2004me}. These short-distance contributions depend on the quark content of the hadron and therefore differ between light, open-heavy, and hidden-heavy sectors. In the present framework, they are absorbed into the physical threshold masses $m_{\rm thr}$, while the exponential growth of excitations above threshold is assumed to be governed universally by long-distance string dynamics characterized by a single universal Hagedorn temperature $T_H$.

In this work, the value $T_H = 323(3)\,$MeV determined from light-hadron thermodynamics\,\cite{Marczenko:2026yme} is used without further adjustment in the heavy-flavor sector.

Fig.~\ref{fig:p_x_ratios} shows partial pressures of open-charm mesons and charmed baryons calculated with the continuous Hagedorn spectrum following Eqs.~\eqref{eq:pres}-\eqref{eq:rho_thresholds}. The solid black line corresponds to the result obtained when the string spectrum is evaluated assuming vanishing charm quark mass, $m_c=0$. In this limit, the entire hadron mass is attributed to string excitations.
This significantly overestimates the lattice QCD results. The mismatch originates from the fact that the string spectrum governs intrinsic excitation energies of the confining flux tube, whereas heavy hadrons carry a large, non-dynamical heavy-quark rest mass. As a consequence, the pressure is strongly enhanced relative to lattice QCD. To correct this, we shift the string excitation energy by the current quark masses, replacing
\begin{equation}\label{eq:rho_shift}
\rho_{\rm str}(m) \;\longrightarrow\; \rho_{\rm str}(m - \sum_q n_q m_q),
\end{equation}
where the sum runs over all quark flavors, with $n_q$ the number of quarks of mass $m_q$ in the hadron. This reflects that the string spectrum governs excitations above the heavy-quark rest mass. It is therefore convenient to introduce the excitation energy,
\begin{equation}\label{eq:Ex}
E \equiv m - \sum_q n_q m_q.
\end{equation}
The separation between endpoint masses and string excitation energy appears naturally in string models with massive endpoints
\cite{Sonnenschein:2014jwa, Sonnenschein:2018fph, Kruczenski:2004me}, as well as in the phenomenology of heavy hadrons in the context of Regge trajectories and in relativistic quark models~(see, e.g.,\,\cite{LaCourse:1988cu, Veseli:1996gy, Jakhad:2025ejc}). In this work, we treat $u$ and $d$ quarks as massless. For the strange quark mass, we use $m_s = 93.5\,$MeV. For the charm quark, we use a PDG benchmark mass, $m_c = 1.273\,$GeV, corresponding to the $\overline{\mathrm{MS}}$ scheme\,\cite{ParticleDataGroup:2024cfk}. The resulting parameter-free spectrum provides quantitative agreement with the partial pressures of open-charm mesons and baryons obtained in lattice QCD (see Fig.~\ref{fig:p_x_ratios}). We note that the case without the shift corresponds to setting all $m_q=0$.

Fluctuations of conserved charges provide a more differential probe of the underlying spectrum than the thermodynamic pressure itself. In particular, fluctuations isolate contributions from distinct quantum-number sectors and are therefore highly sensitive to the mass thresholds and the density of states. To further verify our results, we examine fluctuations of conserved charges involving charm, which are obtained through generalized susceptibilities. The susceptibility of order $n$ is given as a derivative of the thermodynamic pressure with respect to associated chemical potentials\,\cite{Allton:2005gk, Karsch:2010ck},
\begin{equation}
    \hat \chi^{ BQSC}_{jklm}(T,\mu) = \frac{\pp^n \hat P(T, \mu) }{\pp \hat\mu_B^j\pp\hat\mu_Q^k\pp\hat\mu_S^l\pp\hat\mu_C^m}\Bigg|_T,
\end{equation}
such that $j+k+l+m = n$ and $\hat \mu_i = \mu_i / T$. Using the same shifted string density, we compute the ratio of the fourth-order baryon-charm correlator and the charm fluctuations $\hat \chi_{13}^{BC}/\hat \chi_4^C$. This is shown in the middle panel of Fig.~\ref{fig:p_x_ratios}. We find quantitative agreement with lattice QCD over the entire temperature range below the chiral crossover as opposed to the results without the mass shift. Similar results are obtained for ratios involving strangeness, namely $\hat \chi_{13}^{SC}/\hat \chi_4^C$ and $\hat \chi_{22}^{SC}/\hat \chi_4^C$ (see Fig.~\ref{fig:p_x_ratios}). We note that the systematic errors in the computations of $\hat \chi_{22}^{SC}/\hat \chi_4^C$ are large since the extrapolation to the continuum limit is challenging at low temperatures. However, the overall description of LQCD thermodynamics remains quantitatively consistent across channels. This shows that the relevant dynamical variable governing the exponential growth of states is the excitation energy above the current quark mass, rather than the total hadron mass.

\emph{Spectral Abundances.---}
Thermodynamic constraints on the density of states can be confronted directly with hadronic spectroscopy. A direct way to compare the continuous string spectrum with experimentally established hadrons is through the cumulative distribution\,\cite{Broniowski:2000bj, Broniowski:2004yh, Lo:2015cca},
\begin{equation}
    N(m) = \int\limits_0^m \dd m'\, \rho(m'),
\end{equation}
which counts the total number of degrees of freedom below mass $m$. To ensure a consistent comparison, the density $\rho(m')$ must account for the mass shift introduced in Eq.~\eqref{eq:rho_shift}. This representation makes the exponential growth of states manifest and allows for a direct comparison with discrete spectra.

In Fig.~\ref{fig:spectra_bottom} we compare the continuous density of states with the experimental spectra from the PDG\,\cite{ParticleDataGroup:2024cfk} and with relativistic quark-model (QM) predictions\,\cite{Kaczmarek:2025dqt}. When the spectrum is evaluated without the mass shift (not shown in the figure), the resulting cumulative distribution is incompatible with the experimentally established states, reflecting a large contribution of the heavy quark rest mass to the physical hadron masses. In contrast, once the current quark masses are subtracted, the string spectrum aligns remarkably well with the PDG and QM spectra across open-charm mesons and charmed baryons. We note that the QM spectra typically include only states in the valence $q\bar q$ or $qqq$ sector. Once the hadron mass exceeds the lowest two-hadron threshold carrying open heavy flavor (e.g., $M \gtrsim 2M_D$ in the charm sector), the physical states strongly couple to meson–meson or meson–baryon channels and generally appear as resonances in the continuum. Such coupled-channel dynamics is not included in conventional potential models, which therefore typically report only the lowest excitations\,\cite{Chen:2022asf}.

\begin{figure}[t!]
    \centering
    \includegraphics[width=\columnwidth]{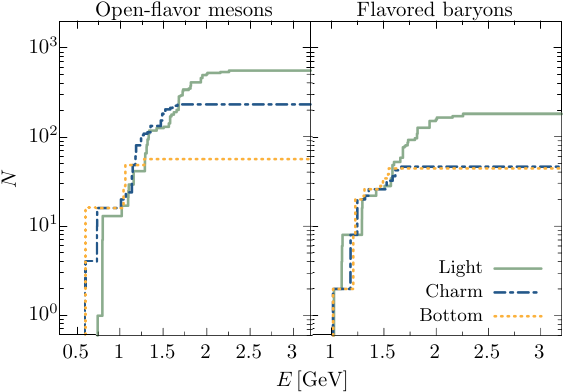}
    \caption{PDG Cumulative spectra across open quark-flavor sectors as a function of the excitation energy $E,$ as introduced in Eq.~\eqref{eq:Ex}. Spectra of mesons and baryons include: $|S|=1$ strange states (Light), $|C|=1$ non-strange (Charm), and $|B'|=1$ non-strange (Bottom) states. The spectra of baryons do not account for antiparticles.
    The light-hadron distributions do not include the Goldstone bosons.}
    \label{fig:universality}
\end{figure}

The excess of baryons predicted by the string spectrum relative to the experimentally established PDG spectrum is consistent with QM studies indicating that the observed charmed-baryon spectrum is incomplete\,\cite{Sharma:2025zhe, Kaczmarek:2025dqt}. The same conclusion is supported by thermodynamic observables: the pressure of charmed baryons and the corresponding fluctuations are consistent with LQCD results, and are enhanced relative to the HRG model with PDG input, signaling additional missing charmed baryons (see Fig.~\ref{fig:p_x_ratios}). These findings may have important consequences for phenomenological applications and for the interpretation of heavy-ion collision data\,\cite{Braun-Munzinger:2023gsd}. 

In the hidden-charm sector, the continuous spectrum also reproduces the overall growth of states. However, we observe that the cumulative string spectrum slightly underestimates the experimentally established PDG states in the intermediate mass region. This deviation is physically understood, as the lowest-lying quarkonia are compact systems dominated by short-distance Coulomb interactions rather than the long-distance linear confining dynamics captured by the Hagedorn spectrum. Given the limited number of experimentally established hidden-charm resonances, small deviations are not unexpected. Nevertheless, the overall exponential trend governed by the same Hagedorn temperature remains consistent with the observed spectrum.

The separation between the quark current mass and the string excitation energy established above (cf.~Eq.~\eqref{eq:rho_shift}) suggests the existence of a universal description of hadronic spectra in terms of the excitation energy across quark flavors. Here, we test this hypothesis by comparing experimental spectra in the bottom sector.

The excitation-energy threshold is fixed separately for each hadronic sector by the lowest physical state, defining $E_{\rm thr} = m_{\rm thr} - \sum_q n_q m_q$ for open-flavor mesons, hidden-flavor mesons, and baryons. This determines the onset of string excitations, which is then assumed to be flavor independent in all hadronic sectors. This procedure yields a current bottom quark mass $m_b \simeq 4.5-4.6\,$GeV, which is consistent with the mass quoted in the PDG~\cite{ParticleDataGroup:2024cfk}.

The experimental spectra from the PDG for charmed and bottomed hadrons, expressed in terms of the excitation energy, are compared in Fig.~\ref{fig:spectra_bottom}. We observe that, within uncertainties, these spectra overlap, indicating that the growth of states is governed by a common excitation-energy scale, which is flavor independent. A similar alignment is observed for the QM spectra. This empirical overlap supports the universality of the excitation spectrum. Consequently, the same Hagedorn density of states, with $T_H$ fixed from the light sector and $E_{\rm thr}$ determined in the charm sector, can be applied to the bottom sector without additional parameters. We note that, after subtracting the current quark masses, small residual differences persist due to uncertainties in the threshold masses entering the density of states. Nevertheless, these differences are numerically small, and the resulting spectra in terms of the excitation energy are practically indistinguishable (see Fig.~\ref{fig:spectra_bottom}).

\emph{Universality Across All Quark Flavors.---}
To compare the spectral growth across different flavor sectors, we compare the experimental cumulative distributions of established resonances in the PDG as a function of the excitation energy, introduced in Eq.~\eqref{eq:Ex}. In the open-flavor meson and flavored baryon sectors, we compare $|S|=1$ strange states, $|C|=1$ non-strange charmed states, and $|B'|=1$ non-strange bottomed states. In these cases, a single strange, charm, or bottom quark is distinguished at the string endpoint. We note that extracting a definitive hidden-flavor spectrum in the light-quark sector is ambiguous due to, e.g., the presence of non-perturbative chiral effects and strong flavor mixing which result in the absence of well-defined quark--antiquark states. We postpone a direct comparison of the hidden-flavor spectrum in the light sector to future work.

The PDG spectra are compared in Fig.~\ref{fig:universality}. We observe that the experimental spectra exhibit a remarkably uniform growth rate. This uniform behavior indicates a universal density of states in terms of the excitation energy. The collapse of these diverse hadronic sectors onto a single characteristic slope provides compelling evidence that the Hagedorn temperature $T_H$ is a fundamental and flavor-independent property of QCD. We note that the observed saturation at higher energies is different for each flavor and is attributed to the current experimental incompleteness of the PDG high-mass spectrum rather than a deviation from universal scaling.

These results demonstrate that the exponential growth of states is controlled by the excitation energy, not the total hadron mass, providing direct evidence that the underlying scale governing the proliferation of all flavors is set by the universal string tension. We note that our framework extends to the top-quark sector, where it defines a corresponding string excitation spectrum governed by the same Hagedorn temperature. However, due to the short lifetime of the top quark, this spectrum is experimentally not well established. Nevertheless, the corresponding excitation spectrum remains well-defined within QCD.

\emph{Summary.---}
We have demonstrated that the open string density of states with a single Hagedorn temperature provides a unified description of hadronic thermodynamics and spectral abundances across quark flavors. In this framework, the exponential growth of the excitation spectrum above threshold is governed by long-distance string dynamics. Once the heavy quark rest mass is separated from the excitation energy, the same Hagedorn temperature that describes light hadrons quantitatively reproduces the thermodynamics of charmed hadrons and, without additional parameters, predicts the bottom spectrum. These results show that the hadronic spectrum of QCD across light and heavy flavors is governed by a universal Hagedorn temperature set by the QCD string tension. Future lattice calculations of heavy-flavor fluctuations and improved measurements of heavy-hadron spectra will provide important tests of this picture. A more precise quantitative study of individual hadronic channels and decay patterns, as well as a systematic comparison with experimental data, is left for future work. Establishing such universality provides a common framework connecting hadronic spectroscopy, lattice QCD thermodynamics, and the description of thermal hadron yields in heavy-ion collisions.

\medskip
\emph{Acknowledgments.---}
The authors acknowledge helpful discussions with Yuki Fujimoto and Chihiro Sasaki. K.R. acknowledges discussions with Peter Braun-Munzinger and  Frithjof Karsch, as well as support from the National Science Centre (NCN), Poland, under OPUS Grant No. 2022/45/B/ST2/01527. K.R. also acknowledges the support of the Polish Ministry of Science and Higher Education.  L. M. thanks partial support
from the Institute for Nuclear Theory, which is funded in part by the INT’s U.S. Department of Energy grant No. DE-FG02-00ER4113.

\appendix

\section{\large \bf End Matter}
\section{Construction of the string spectrum}

In the present framework, hadrons are described as open relativistic strings whose endpoints carry quarks\,\cite{Nambu:1974zg, Isgur:1984bm}. The long-distance confining interaction in QCD is characterized by a linear potential, motivating the identification of mesons and baryons with color flux tubes of tension $\sigma$. Excitations of this flux tube generate an exponentially growing density of states governed by the Hagedorn temperature $T_H$\,\cite{Hagedorn:1965st, Hagedorn:1971mc}, which depends only on the string tension and not on the flavor of the endpoint quarks\,\cite{Green_Schwarz_Witten_2012}. Flavor quantum numbers enter through the endpoints, while the physical ground-state masses set the excitation thresholds. 

In constructing the string spectrum, heavy-quark spin symmetry (HQSS) is treated as exact at leading order. In the heavy-quark limit, the spin of the heavy quark decouples from the light degrees of freedom, and hadrons organize into nearly degenerate spin multiplets. Members of a given multiplet correspond to the same underlying string configuration and therefore share a common excitation threshold, fixed by the lowest mass in the multiplet. Hyperfine splittings arise from short-distance spin--spin interactions suppressed by $\mathcal{O}(m_Q^{-1})$ and do not generate independent string excitations.

Mesons are constructed as open strings with a quark at one endpoint and an antiquark at the other. Open heavy-flavor mesons correspond to configurations $Q\bar q$ and $\bar Q q$, where $Q=c,b$ and $q=u,d,s$. The excitation threshold is fixed by the lowest pseudoscalar states $D$, $D_s$ for charm and $B$, $B_s$ for bottom.

Hidden heavy-flavor mesons correspond to strings with heavy quark and antiquark at both endpoints, $Q\bar Q$. For such systems, the heavy quarks are approximately nonrelativistic, allowing an effective-field-theory description in terms of nonrelativistic QCD\,\cite{Bodwin:1994jh, Brambilla:2004jw}. The lowest physical states are the pseudoscalars $\eta_c$, $\eta_b$, and their vector partners $J/\psi$, $\Upsilon$, which form a heavy-quark spin multiplet corresponding to a common string ground state.

Baryons are modeled as open strings with a quark attached to one endpoint and a diquark attached to the other~\cite{Fujimoto:2025trl}. This construction is motivated by the color decomposition $\mathbf{3}\otimes\mathbf{3}=\bar{\mathbf{3}}\oplus\mathbf{6}$, where the antisymmetric $\bar{\mathbf{3}}$ channel is attractive. Quark pairs in this channel form correlated diquark degrees of freedom. We denote antisymmetric (scalar) diquarks by $[qq]$ with spin $s_{qq}=0$, and symmetric (axial) diquarks by $\{qq\}$ with spin $s_{qq}=1$.

Singly heavy baryons such as $\Lambda_Q$ and $\Xi_Q$ contain scalar diquarks, whereas $\Sigma_Q$, $\Xi'_Q$, and $\Omega_Q$ contain axial diquarks. Scalar diquarks with $j_\ell=0$ yield a single $J=\tfrac{1}{2}$ state, while axial diquarks with $j_\ell=1$ produce nearly degenerate doublets with $J=\tfrac{1}{2},\tfrac{3}{2}$. For doubly heavy baryons, the two heavy quarks form a compact diquark $\{QQ\}$ in the attractive $\bar{\mathbf{3}}$ channel. By Pauli symmetry, identical heavy quarks require a symmetric spin wavefunction, implying an axial configuration with $s_{QQ}=1$. The system can therefore be described as a light quark attached to a heavy axial diquark, with the light degrees of freedom carrying $j_\ell=\tfrac{1}{2}$ and a resulting $J=\tfrac{1}{2},\tfrac{3}{2}$ doublet. Triply heavy baryons can be described in the same quark--diquark framework as a heavy quark attached to a heavy diquark, $Q\{QQ\}$, yielding a single $J=\tfrac{3}{2}$ ground state fixed by the symmetry of identical quarks.

The string excitation spectrum depends only on the excitation energy above the physical threshold. The masses of mesons and singly-flavored baryons are taken from PDG~\cite{ParticleDataGroup:2024cfk}. The masses of doubly and triply charmed thresholds are taken from QM predictions~\cite{Ebert:2002ig, Zhou:2025fpp, Chen:2022asf}. Owing to their large masses, the thermal contribution of doubly and triply heavy baryons is exponentially suppressed at temperatures $T \lesssim T_c$ and is therefore negligible for the thermodynamic analysis performed in this work.

The hadronic thresholds entering Eq.~\eqref{eq:rho_thresholds} are grouped according to mesonic and baryonic sectors, with the corresponding states, quantum numbers, and degeneracy factors listed in Tables~\ref{tab:mes} and~\ref{tab:bar1}. Degeneracy factors $g$ account for spin, isospin, assuming exact HQSS spin-multiplet grouping.

\begin{table}[t!]
\centering
\begin{tabular}{c c || c c c | c c | c}
\hline
\multicolumn{2}{c}{} &\multicolumn{3}{c}{} & \multicolumn{2}{|c|}{Mass\,[GeV]} & \\
State & String & $J$ & $I$ & $S$ & $Q=c$ & $Q=b$ & $g$\\
\hline\hline
$D$, $B$ & $Q\bar l,\,\bar Q l$ & $0$ & $1/2$ & $0$  & 1.867 & 5.279 & 16\\
$D_s$, $B_s$ & $Q\bar s,\, \bar Q s$ & $0$ & $0$   & $\pm1$ & 1.968 & 5.367 & 8\\
\hline
$\eta_c$, $\eta_b$ & $Q\bar Q$ & $0$ & $0$   & $0$ & 2.984 & 9.398 & 4 \\
\hline
\end{tabular}
\caption{Open- and hidden-flavor thresholds for  mesons with heavy $Q=c,b$, light $l=u,d$ and strange $s$ quarks. $J$, $I$, $S$, and $g$ refer to spin, isospin, strangeness, and degeneracy, respectively.}
\label{tab:mes}
\end{table}

\begin{table}[!t]
\centering
\begin{tabular}{c c || c c c | c c | c}
\hline
\multicolumn{2}{c||}{} &\multicolumn{3}{c}{} & \multicolumn{2}{|c|}{Mass\,[GeV]} & \\
State & String & $J$ & $I$ & $S$ & $Q=c$ & $Q=b$ & $g$ \\
\hline\hline
$\Lambda_Q$ & $Q[ll]$ 
& $1/2$ & $0$ & $0$ & 2.286 & 5.620 & 2 \\
$\Sigma_Q$ & $Q\{ll\}$ 
& $1/2,\,3/2$ & $1$ & $0$ & 2.453 & 5.811 & 18 \\
\hline
$\Xi_Q$ & $Q[ls]$ 
& $1/2$ & $1/2$ & $-1$ & 2.468 & 5.793 & 4 \\
$\Xi'_Q$ & $Q\{ls\}$ 
& $1/2,\,3/2$ & $1/2$ & $-1$ & 2.576 & 5.935 & 12 \\
\hline
$\Omega_Q$ & $Q\{ss\}$ 
& $1/2,\,3/2$ & $0$ & $-2$ & 2.695 & 6.046 & 6 \\
\hline
$\Xi_{QQ}$ & $l\{QQ\}$ 
& $1/2,\,3/2$ & $1/2$ & $0$ & 3.519 & -- & 12 \\
\hline
$\Omega_{QQ}$ & $s\{QQ\}$ 
& $1/2,\,3/2$ & $0$ & $-1$ & 3.778 & -- & 6 \\
\hline
$\Omega_{QQQ}$ & $Q\{QQ\}$ 
& $3/2$ & $0$ & $0$ & 4.800 & -- & 4\\
\hline
\end{tabular}
\caption{Heavy baryon thresholds with heavy $Q=c,b$ and light $l=u,d$, $s$ quarks. Square brackets denote antisymmetric (scalar) diquarks and curly brackets symmetric (axial) diquarks. $J$, $I$, $S$, and $g$ denote spin, isospin, strangeness, and degeneracy, respectively.}
\label{tab:bar1}
\end{table}

\bibliographystyle{apsrev4-2}
\bibliography{sample}

\end{document}